\documentclass[twocolumn,pre,a4paper,showpacs]{revtex4}
\usepackage{amsbsy}
\usepackage{amssymb}
\usepackage[dvips]{graphicx}

\begin{document}

\title{Distribution of essential interactions in model 
gene regulatory networks under mutation-selection balance}

\author{Z. Burda $^1$, A. Krzywicki $^2$, 
O.C. Martin $^{3,4}$ and M. Zagorski $^1$} 

\affiliation{$^1$ Marian Smoluchowski Institute of Physics
and Mark Kac Complex Systems Research Centre, Jagellonian University,
Reymonta 4, 30-059 Krakow, Poland\\
$^2$ Univ Paris-Sud, LPT ; CNRS,
UMR8627, Orsay, F-91405, France.\\
$^3$ Univ Paris-Sud, LPTMS ; CNRS, UMR 8626, 
F-91405, Orsay, France\\ 
$^4$ INRA, CNRS, UMR 0320 / UMR 8120 G\'en\'etique V\'eg\'etale, 
F-91190 Gif-sur-Yvette, France}

\date{\today}

\begin{abstract}
Gene regulatory networks typically have low in-degrees, whereby 
any given gene is regulated by few of the genes in the network.
They also tend to have broad distributions for the out-degree.
What mechanisms
might be responsible for these degree distributions?
Starting with an accepted framework of the 
binding of transcription factors to DNA, we consider
a simple model of gene regulatory dynamics. There,
we show that selection for a target 
expression pattern leads to the
emergence of minimum connectivities compatible with
the selective constraint. 
As a consequence, these gene networks have low in-degree, 
and ``functionality'' is parsimonious, \emph{i.e.}, 
is concentrated on a sparse number
of interactions as measured for instance by their essentiality. 
Furthermore, we find that mutations of the transcription 
factors drive the networks to have broad out-degrees.
Finally, these classes of models are evolvable,
\emph{i.e.}, significantly different genotypes 
can emerge gradually under mutation-selection balance.
\end{abstract}

\vspace{2pc}

\maketitle

\section{Introduction}
It is by now well established that complex organisms
typically do not have many more genes than less complex ones. 
Because of this, the paradigm for thinking about biological complexity 
has shifted from the number of genes to the way they work together: 
higher complexities might be associated with
a greater proportion of regulatory genes. In particular,
there are strong indications in eukaryotes and prokaryotes
that for increasing genome size the number of 
regulatory genes grows faster than linearly in the total
number of genes~\cite{Nimwegen2003,BabuLuscombe2004}. Hence it is 
appropriate to consider biological complexity in the framework of
interaction networks. This shift from components
to the associated interactions
has received increasing attention in many scientific communities,
with applications ranging from network biology to sociology.
The relevance of this conceptual framework for biology has 
been repeatedly emphasized 
and has benefited from inputs from other fields and from
statistical physics in 
particular~\cite{AlbertBarabasi2002,BarabasiOltvai2004}.
We will therefore freely use the \emph{network} terminology, 
refering to nodes, their degrees, 
distinguishing between in and out-degrees etc.

From studies that strive to unravel gene regulatory networks (GRN), 
several qualitative properties transpire:
(i) a given gene is generally influenced by a ``small'' number of
other genes (low in-degree of the network of interactions when
compared to the largest possible 
degree~\cite{ThieffryHuerta1998,BalazsiHeath2008}); 
(ii) some genes are very pleiotropic (the out-degree of some
nodes of the network can be high, leading to degree distribution
with fat tails or even possibly scale-free 
behavior~\cite{BarabasiOltvai2004,BalazsiHeath2008}); (iii) GRN seem
to be robust to change (e.g., to environmental
fluctuations or to mutations), a feature
that is also found at many other levels of biological 
organisation~\cite{BornholdtSneppen2000,EdlundAdami2004,
WagnerBook2005,ChavesAlbert2005}.
A simple way to build robustness into a network is to have rather 
dense connections, effectively incorporating redundancy,
either locally or globally. Furthermore, the number
of networks having $m$ interactions grows very quickly with $m$. Thus
when modeling GRN, the \emph{network} realizations that perform a
given regulatory function are dominantly of very high degree.
However this is not the case experimentally, at least with respect
to the in-degree, and so models so far have had to 
build-in limitations to the accessible
connectivities~\cite{KauffmanBook1993,AldanaCluzel2003,KauffmanPeterson2004}.
In this work we show that such shortcomings of models 
are overcome if one takes into account the known mechanisms 
underlying genetic interactions: gene regulation is mediated
via the molecular recognition
of DNA motifs by transcription factors, and this leads
to biophysical constraints on interaction strengths. Within this 
relatively realistic framework,
we shall see that networks under mutation-selection balance:
(i) are driven to be parsimonious
(the essential interactions are sparse) for the in-degree;
(ii) can have broad distributions for the out-degree 
which is unconstrained;
(iii) can evolve to very different realizations while preserving
their function.

We begin by explaining the mechanisms incorporated into
the model, in particular the determinants of the
interactions. We follow standard 
practice~\cite{HippelBerg1986,BergHippel1987,GerlandMoroz2002} when
modeling interactions between DNA binding sites
and transcription factors (TF): the affinity
depends on the mismatch between two character strings.
We also specify how gene expression dynamics depend
on these interactions and what 
``function'' the networks must implement.
The main difference between our approach and previous work
is the introduction of a molecular definition of 
genotypes; this more realistic setting means that
mutations are no longer ad-hoc and interestingly this difference leads
to all the generic properties listed above.
In fact, we keep our model as simple as possible to drive home 
the fact that all these properties emerge quite inevitably once such a
setting is used. Clearly, our choice of a ``minimal'' model 
means that we must focus on qualitative
aspects of the problem without attempting to reproduce 
specific experimental data.

After giving the general framework, we present some of the
mathematical and computational tools
we use to analyze the model.
\emph{First}, we demonstrate
that selection (the constraint of having a 
given function or ``phenotype'') 
makes the networks'
connectivity be close to minimal,
\emph{i.e.}, networks are as sparse as they can be
subject to maintaining their function. 
The ``spontaneous'' appearance of sparsity is due to the fact
only a tiny (negligible) fraction of functional genotypes correspond to 
non-sparse networks. This \emph{entropic} property is in strong
contrast to what happens in models formulated in terms of networks (rather
than in terms of microscopic genotypes).
A consequence of this sparsity is that the genotypes 
are quite robust to 
mutations~\cite{NimwegenCrutchfield1999,WagnerBook2005}: only those 
few binding sites that
are ``effectively'' used are fragile, mutations of the
other (little used) binding sites have almost no effect.
Thus robustness to mutational changes is very high for
most binding sites while the ``essential'' interactions
have much lower robustness; robustness is heterogeneously distributed
in the network. \emph{Second}, we find that the network
out-degree is unconstrained, but that
under evolutionary pressure coming from mutation-selection balance,
broad out-degree distributions are favored.
Implications of our work are developed
in the discussion; in particular, a consequence
is that redundant interactions are 
rapidly eliminated under evolution if no new
function arises which might change the selection pressure.

\section{The model}

\subsection{Model genealogy}

Gene regulatory networks play an essential role in cellular dynamics.
A question of major interest is how cells 
maintain their integrity; this includes
stabilizing gene expression in the presence of 
environmental or genetic perturbations~\cite{WaddingtonBook1957},
actively maintaining oscillatory dynamics as in circadian
clocks~\cite{GoldbeterBook1997}, or responding 
transiently and in a timely way
to external signals~\cite{AlonBook2006}.
In the majority of studies tackling these issues theoretically, 
one attempts to adjust model or kinetic 
parameters in gene circuits to obtain 
a GRN with the desired properties. Our focus here is different:
we ask how functionality constrains network structure. Answering
this requires considering the ``space'' of all functional GRN
and determining the generic properties arising in this space. 
In particular, we shall be interested in the topological features
and the robustness of these GRN.

The use of a space of GRN to propose ``design principles'' of gene regulation has
been exploited previously by several 
groups~\cite{KauffmanBook1993,WagnerBook2005}.
Such a paradigm enables one to put forward certain generic 
behaviors, and it is often complementary to other approaches. Perhaps 
the most widespread such
framework was made popular by Stuart 
Kauffman (see~\cite{KauffmanBook1993} 
and references therein) using random boolean networks.
Each node of the network is associated with a gene and its 
expression level which, by assumption, is a \emph{boolean}
variable (1/0 for on/off) representing the possible two extreme expression 
levels of the gene. The gene expression dynamics at each node
is taken as a random boolean function of its inputs; these inputs are
the gene expression levels of the nodes it is directly 
connected to. Gene expression 
patterns or ``phenotypes'' are then represented by $N$-dimensional
vectors with boolean components.
Among other subjects, this framework was used to  
study a number of generic properties arising in this ensemble
such as robustness to noise and mutations, 
evolutionary paths in classes of fitness landscapes and
consequences for network topology. 

Kauffman's approach led to the creation of
entire familes of boolean models.
Further steps were made by people trying to get additional insight 
by putting more knowledge into the boolean functions. Many
works did this by focusing on ``threshold'' networks
which had already been widely used in neural network modeling.
There, the random boolean functions for each node are
replaced by an ``integrate and fire'' type relation.
To this end, one introduces an $N\times N$ matrix of inter-gene 
interactions~\cite{Wagner1996} where $W_{ij}$ is the strength of 
the interaction of the $j$-th gene on the $i$-th one. 
(Notice that this
matrix can be regarded as representing a directed weighted graph.)
Typically these models remain boolean, so that 
the ``output'' is ``on'' if and only if the incoming
(integrated) signal is above a threshold; the framework can
then be considered as a refinement of Kauffman's. Models of 
this class have also been very popular, and have been used to study
robustness, evolvability and other ``generic''
properties~\cite{BornholdtRohlf2000,LiLong2004,AzevedoLohaus2006,
CilibertiMartin2007a, Leclerc2008}.
However, their main shortcoming is that the
$W_{ij}$ weights are introduced in an ad-hoc way, with the consequence that 
any evolution of these networks is also arbitrary. In particular,
one can deplore that these weights are in no way connected to 
known features of the underlying microscopic processes giving
rise to genetic interactions.

Although we conceptually borrow much from 
the work of our predecessors, we also depart
considerably from these earlier studies: we abandon the boolean
approximation, we adapt known thermodynamical considerations to
formulate our transcriptional dynamics, and we make use of the present
knowledge of molecular processes to introduce in a 
realistic fashion interaction weights $W_{ij}$ and their dependence
on mutations.

\subsection{Genotypes}

As already mentioned, GRNs are often represented by a directed weighted 
graph, \emph{i.e.}, by an $N \times N$ matrix 
$\{W_{ij}\}$, where $N$ is the number
of genes belonging to the GRN. 
The $\{W_{ij}\}$ matrix can \emph{a priori} be quite arbitrary.
In complete generality, the products of the 
$N$ genes can have regulatory influences on the same 
set of genes (retroaction) and possibly also 
have some ``down-stream'' consequences on other genes that
do not code for TF. 
However the consequences of these last 
effects can be ignored for our purposes 
since they lead to no feedback on the $N$ ``core'' genes. 
The focus of all this paper is thus on such a core GRN, containing
only TF coding genes. Note that these restricted networks
have similar statistical properties to the unrestricted ones,
namely sparsity, very low values for the
in-degree, and a fat tail for the out-degree 
distribution~\cite{BalazsiHeath2008}.

Since it is the affinity between the 
transcription factor $j$ and its binding site
(a DNA sequence) in the 
regulatory region of gene $i$ that microscopically controls the
level of transcription, we want $W_{ij}$ to be a measure of this affinity.
An affinity depends
on the binding free energy which itself determines
the frequency with which these two molecules will
be bound rather than unbound. 
Following standard practice, we represent each TF as well 
as each binding site by a character 
string of length $L$; we also impose these characters
to belong to a 4 letter alphabet in direct analogy with the
four bases of DNA. The list of these strings then
defines the \emph{molecular genotype} of the system under consideration.

How can one connect this molecular genotype to a set of interactions
$W_{ij}$ in the GRN? Building on the well known work of Berg and von 
Hippel~\cite{HippelBerg1986}, we assume for 
the sake of simplicity that the free energy of 
one TF molecule bound to its target is, up to an additive
constant, equal to $\varepsilon d_{ij}$, 
where $d_{ij}$ is the number of \emph{mismatches} between the strings 
representing the $j$-th 
TF and its binding site in the regulatory region of gene $i$.
The parameter $\varepsilon$ is the penalty for each 
mismatch, \emph{i.e.}, the contribution to binding 
free energy in units of $k_B T$
where $k_B$ is the Boltzmann constant and $T$ is the
temperature in degrees Kelvin. Experimentally, 
$\varepsilon$ is inferred to have values between one and three if one thinks
of each base pair of the DNA as being represented by one
character~\cite{SaraiTakeda1989,StormoFields1998,BulykJohnson2002}.
Note also that comparing to the typical number of base pairs 
found in experimentally studied binding sites leads to $10 \le L \le 15$. 
Given this framework, we now \emph{define} the ``interaction
strengths'' $W_{ij}$ arising in such a GRN via the Boltzmann factor
\begin{equation}
W_{ij} = e^{-\varepsilon d_{ij}} / Z  
\label{eq:Wij}
\end{equation}
where $Z$ is a normalization (actually a partition function).
If there were just one transcription factor molecule of type $j$, 
$W_{ij}$ would be the probability to find that molecule bound
onto its binding site in the regulatory region of gene $i$.
Gerland et al.~\cite{GerlandMoroz2002} have shown
that the term $Z$ in Eq.~(\ref{eq:Wij}) is in practice
close to 1 and that the probability of finding a particular
transcription factor molecule bound at a given binding site is low.

If we consider the space of all genotypes, that is all possible
character strings in our model with equal probability, the 
\emph{a priori} distribution (coming from random strings) of the 
mismatch $d$ for any given pair $(i,j)$ is binomial:
\begin{equation}
p(d) = {L\choose d} (1/4)^{L-d} (3/4)^{d} \; .
\label{eq:binom}
\end{equation}
Then for the biologically realistic values of $L$ mentioned above, using
for instance Stirling's formula, we have
\begin{equation}
p(d) \sim \frac{3^d}{d \, !} 
\frac{L^d}{4^L} \ll 1  \; \; \text{when} \; \; \; d \ll L \; .
\label{eq:smallness}
\end{equation}
Hence small mismatches are very improbable, 
a fact that will be of utmost importance 
later on. 

\subsection{Occupation of the binding sites}

Suppose, that there are $n_j$ TF molecules of type $j$ that can bind to a
site in gene $i$'s regulatory region; given that this site can be occupied
only by one TF molecule at a time, it is necessary to take into account
the possible competition. Using the fact that $Z$ in
Eq.~(\ref{eq:Wij}) is close to 1, it is possible to approximate
the occupation probability of the binding
site by~\cite{GerlandMoroz2002}:
\begin{equation}
P_{ij} = \frac{1}{1 + 1/(n_j W_{ij})}  \; .
\label{eq:probOccupy}
\end{equation}

In the following, we denote by $n$ the maximum number of 
TF molecules that can arise when a gene is fully ``on'', taking
for simplicity this maximum to be independent of $j$.
Biologically, $n$ is known to have a wide range, going from 
of order unity to many thousands.
The lower value comes from some measurements of the multiplicity 
of transcripts~\cite{GoldingPaulsson2005,BecskeiKaufmann2005},
while the higher value
comes from other measurements of the numbers 
of transcription factor molecules~\cite{ElfLi2007}. For our study,
we consider several values 
of $n$, reporting on the range $10 \le n \le 10^4$.
We also considered smaller values for $n$ and found little change, but
when $n$ is of order 1, our ``mean field'' framework 
which neglects fluctuations can
no longer be justified.

Setting $n$ to be $j$-independent is a 
rather strong assumption. It is made here
for the sake of simplicity, to avoid a 
proliferation of free parameters. 
The implication is that we abandon any 
attempt to introduce external control
factors. For example, it is known that the 
probability of a CAP molecule to
attach to DNA and recruit a polymerase depends 
not only on the number of these
molecules but on the concentration of 
glucose~\cite{BusbySavery2007}. By the same token,
including co-factors or repressors or modeling more
refined feedback circuits is beyond the scope of this paper. 

\subsection{Phenotypes}

Call $S_j(t)$ the normalized level of gene $j$'s products at time $t$, 
corresponding to a total number of TF molecules~\cite{PaboSauer1992}
equal to $nS_j(t)$. As already mentioned, we abandon the boolean
approximation whereby a gene's expression level is either
at its minimum or its maximum. Instead, we allow all intermediate
values making $S_j(t)$ a
continuous variable. In practice, we take 
all the $S_j(t)$ to lie in the interval $[0,1]$, corresponding to 
a minimum number of TF molecules of a given type equal to 0 and 
a maximum number equal to $n$.
A short remark is in order for readers 
nourished with boolean models to avoid confusion: of 
course, instantaneously, a gene is either being transcribed (on) or not
(off). However, these transcripts have to be translated into proteins
(TFs) and these transcription factors will then have a certain half-life.
Thus, it is necessary to distinguish the ``effective'' or 
average level of
expression of a gene $j$ (here the number 
of TF molecules of type $j$ present)
from the instantaneous transcription rate. 
In a model like this one, only
the effective expression level is relevant.

At any given time, there is a vector of effective expression levels.
In the next subsection we will define a deterministic dynamical system to
model the time evolution of this vector. After possible transient behavior, 
at large times the set of expression levels $\{ S_j \}_{j=1,\ldots N}$ 
may go to a (time-independent) steady state; we 
refer to this final vector as the ``phenotype'' of the GRN.
It is also possible for the dynamics to go into a cycle
(periodic behavior), but we shall not consider those
GRNs in our analysis. Note that which case arises may 
depend on the initial expression levels. 

GRNs enable the homeostasis of gene expression, allow 
for a response to a stimulus,
or realize a new function such as cellular differentiation.
In this work we consider that the ``function'' of our GRN 
is to bring the gene expression to a given pattern and maintain it there.
This choice~\cite{Wagner1996} is motivated from cases arising in
early embryo development; there, the initial 
expression levels, $\{ S_j^{(initial)} \}_{j=1,\ldots N}$, are given
(inherited during the formation of the egg).
Then a network will ``perform the desired function''
if and only if, starting with the initial expression pattern,
the vector of expression levels converges to a steady state
which is close enough to a ``target'' pattern, which again must
be given \emph{a priori}. We will denote these target levels
by $\{ S_j^{(target)} \}_{j=1,\ldots N}$.
Hereafter we loosely refer to a network as ``viable'' if it
statisfies well this functional property, \emph{i.e.}, 
if its phenotype is sufficiently close to the target one.

\subsection{\bf Expression pattern dynamics}

The dynamics of the $S_j(t)$ takes place 
on biochemical time scales (typically 
minutes). To model these dynamics, we 
consider that the normalized level
$S_i(t+1)$ of the $i$-th gene at time $t+1$
is strongly associated with its transcription rate at time $t$.
Clearly, that transcription rate depends on the degree to which the gene's
regulatory region is occupied by TFs. 
To find the probability that a TF of type $j$ is bound to
gene $i$'s regulatory region,
we use Eq.~(\ref{eq:probOccupy}) but where the
term $n_j$ is replaced by $n S_j(t)$:
\begin{equation}
P_{ij} = 1/(1 + h/W_{ij} S_j(t))  \; .
\label{eq:dynamicsBis}
\end{equation}
Here we have introduced $h=1/n$ which plays the role 
of an effective threshold for the 
action of $W_{ij}$. This helps in comparing 
our results with those obtained 
with threshold models (including those from ref.~\cite{Burda2009}).

To keep the framework simple,
we shall suppose that gene $i$'s transcription is ``on'' whenever \emph{at least}
one TF is bound within its regulatory region and otherwise it is ``off''. 
This is reminiscent of an ``OR'' logical gate whereby the output
is on if and only if at least one of the inputs is on.
The (normalized) mean expression level of a gene is then identified with
the probability that transcription is on. 
Taking the TF occupancies in the regulatory region
to be statistically independent, we then have
\begin{equation}
S_i(t+1) = 1 - \prod_j (1 - P_{ij})
\label{eq:dynamics}
\end{equation}
Eqs.~(\ref{eq:dynamics}) and (\ref{eq:dynamicsBis})
define our discrete time transcriptional dynamics.

Two qualitative remarks are in order.
First, we have taken the occupation probabilities of the different
TFs to be independent; 
our framework thus ignores collective binding effects and in particular
interactions between TF. 
Second, we have restricted the transcriptional logic to be of the
``OR'' type. Our TF thus act only as enhancers, 
never as repressors, and they do not cooperate using more
complicated logic~\cite{BuchlerGerland2003}. Since repressors are widespread
in GRN, we have looked into generalizations of our model: we allowed
repressor interactions while at the same time 
forcing multiple target expression patterns for a GRN to be viable.
The resulting GRN are still sparse and their out-degree distribution
still has a fat tail. 
However, since this paper's goal is to show how sparsity
and high out-degrees emerge within minimal hypotheses,
we postpone presenting these generalizations to a later work.

One can also ask whether our model is structurally robust. Any
real expression dynamics will include fluctuations in the number
of TF molecules~\cite{BecskeiKaufmann2005,BlakeBalazsi2006,BarkaiShilo2007}
whereas our $S_i(t)$ dynamics are mean-field-like
and thus neglect fluctuations. In general such an approximation can
be justified when the number of molecules involved is large,
\emph{i.e.}, in our case when $n$ is large. However, it is rarely
clear \emph{a priori} what is ``large enough''. Thus we have investigated
the effects of adding noise to Eq.~(\ref{eq:dynamics}), letting
the number of molecules $n S_i(t)$ become a Poisson random variable
whose mean is given by that equation. Interestingly, 
none of the properties of the resulting GRN are significantly modified, even
when going down to $n=10$. This can be understood qualitatively 
by considering the robustness of Eq.~(\ref{eq:dynamicsBis}). 
$P_{ij}$ in that equation will be affected by noise in $S_j(t)$,
but much less than by the modification of $W_{ij}$ induced by changing 
the mismatch $d_{ij}$ by one unit. One can then expect that high
level of noise in the expression dynamics is unable to compensate
the relevant mismatches present in the genotype.

\section{Implementation}

\begin{figure}
\includegraphics[width=8cm]{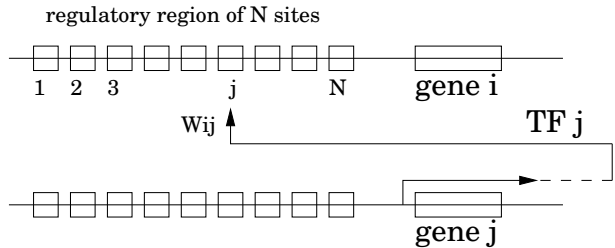}
\caption{\label{fig:model} Schematic representation of the regulatory region
of gene $i$: there are $N$ binding sites, each labeled by an index
$j$ ($1 \le j \le N$). Represented is the interaction $W_{ij}$ 
mediated by the binding of TF $j$  to the $j$'th site of that region. 
The binding affinities depend on the mismatch between the string of 
length $L$ representing the TF and that representing the DNA of the 
corresponding binding site.}
\end{figure}

To minimize the complexity of the model, we consider that each gene's
regulatory region consists of $N$ putative binding sites,
one for each of the $N$ types of TFs as illustrated
in Fig.~\ref{fig:model}. If gene's $j$ normalized 
expression level is $S_j$, it 
will produce a certain number $nS_j$ of TF molecules of type $j$.

The genotype of one of our networks consists of many strings of length $L$
containing characters from a four letter alphabet.
Explicitly, for each of the $N$ genes, 
one string encodes the TF and $N$ strings encode the DNA sequences
of the different binding sites in the regulatory region of that gene. There
is thus a total of
$N(N+1)L$ characters specifying each genotype. 
This in fact defines the (discrete)
space of \emph{all} genotypes. However we are not interested in all genotypes, 
we want those that have ``good'' phenotypes. This leads us to define
quantitatively a fitness for each genotype as follows.

Let us first define a ``distance'' between two 
phenotypes ${\mathbf S}$ and ${\mathbf S'}$ using the differences in
expression levels for each gene:
\begin{equation}
D({\mathbf S}, {\mathbf S'})=\sum_i \mid S_i-S_i'\mid \; .
\label{eq:distance}
\end{equation}
Algorithmically, we consider that the dynamical 
process Eq.~(\ref{eq:dynamics}) has reached 
a fixed point if
$D({\mathbf S}(t+1), {\mathbf S}(t)) < 10^{-8}$. 
Given a phenotype, we take its fitness to be 
\begin{equation} 
F({\mathbf S})=\exp{(-f D({\mathbf S},{\mathbf S}^{(target)}))}
\label{eq:fitness}
\end{equation} 
where $f$ is a control parameter. A simple 
argument indicates that $f$ should be quite
large: the maximum expression of gene $i$ 
(\emph{i.e.}, $S_i=1$) corresponds to the production of 
$n + O(\sqrt{n})$ TF molecules if one allows for Poisson noise. 
(This noise is unavoidable: diffusion as well as other processes
necessarily lead to statistical fluctuations.) The relative 
fluctuation in $S_i$ is 
therefore $O(1/\sqrt{n})$. Such a fluctuation must be innocuous 
for the cell and
must not generate significant loss in fitness.
Hence, one should roughly have 
$f /\sqrt{n}$ of order 1 or less, and thus $f = O(\sqrt{n})$ or less. 
In our simulations we use $f=20$; the results depend only very weakly 
on this choice provided $f$ is in the range 10 to 100.

Given this definition of fitness, we can sample the fitness landscape
of our system by Monte Carlo importance sampling. The goal is to bypass
genotypes of low fitness and to focus instead
on those genotypes whose phenotype is close to the target one.
A standard approach to do this is to use the fitness
as the measure to be used in the sampling: each genotype will appear with
a probability proportional to its fitness. We do this by the
Metropolis algorithm, producing a (biased) random walk in
the fitness landscape which visits successive
genotypes. At each step of the algorithm, we
go to a neighboring genotype (in practice this is
done by changing one character in the strings defining
the genotype). Then the phenotype and fitness of this
modified genotype is determined; the modified genotype is
accepted or not, and the process is repeated.
The acceptance procedure uses the 
Metropolis rule: if the fitness has increased,
the modified genotype is accepted; if the fitness has decreased,
the modified genotype is accepted with a probability
given by the ratio of the new and old fitnesses.
After many such steps, the algorithm will
sample the space with the specified measure.
In the next section, we shall consider the case
where the TF character strings are fixed; the sampling of genotypes
then restricts the modifications to arise only in
the regulatory regions. In that case, we use
``sweeps'', a sweep being a succession of $LN^2$ 
attempted modifications.

For simplicity, we shall choose the 
$\{ S_j^{(initial)} \}_{j=1,\ldots N}$
and 
$\{ S_j^{(target)} \}_{j=1,\ldots N}$
to be ``on'' (1) or ``off'' (0). 
If $S^{(initial)}_{i=1,\dots,N}$ and 
$S^{(target)}_{i=1,\dots,N}$ are drawn at random,
the number of components set to 1 will be approximately 
equal to that set to 0 when $N$ is large. 
For the results presented here, 
these numbers are set to $N/2$ exactly as this reduces
finite size effects in $N$. One can of course also
use other choices for these numbers. To check the robustness
of our model's properties, we have examined how 
the network connectivity is affected
when we take ${\mathbf S}^{(target)}$ to have $N/4$ or
$3N/4$ components set to 1 instead of the $N/2$ 
used throughout this paper. Interestingly, in all cases, 
we find that the same qualitative properties emerge, namely
that for each gene that is ``on'' in ${\mathbf S}^{(target)}$
there is one incoming essential interaction (as defined later),
while the out-degree distribution of the network has a fat tail.
Coming back to our present choice motivated by drawing
the expression vectors at random,
we see that because of the permutation 
symmetry of the model, one can
always swap the indices so that $S^{(initial)}_i = 1$ 
for $i \leq N/2$ and 0 otherwise;
furthermore we also impose without loss of generality
$ S^{(target)}_i=1$ for $N/4 < i \leq 3N/4$ and 0 otherwise.
Notice that $\sum_i S^{(initial)}_i = \sum_i S^{(target)}_i = N/2$
and $\sum_i S^{(initial)}_i S^{(target)}_i = N/4$.

\section{Binding site mutations and sparsity}

\subsection{The emergence of a viable genotype}

We start by considering the TF encodings as
given and fixed. The justification for this comes 
from the fact that because TF are typically 
pleiotropic, they are subject to strong stabilizing selection;
they are thus generally thought to 
evolve slowly, while DNA regulatory regions 
typically have a high level of polymorphism 
and evolve more quickly~\cite{HuynenBork1998}. In 
this and the next subsection, we adopt, as 
a first approximation, the assumption that the genes (and 
thus the TF they code) are fixed whereas 
the strings of characters representing the 
DNA binding sites are unconstrained. A 
consideration of the effects of TF mutations 
is postponed to the following section.

As already mentioned, the genotype determines the list of 
weights $W_{ij}$, corresponding to a 
weighted oriented graph. The first problem we 
address is the emergence of a viable genotype starting from an entirely
random genotype. This can also be done by taking the 
$W_{ij}$'s using mismatches generated from the binomial distribution
Eq.~(\ref{eq:binom}). Random genotypes 
will produce very poor (unfit) phenotypes,
but if the system is allowed to evolve under a selection
pressure on the fitness, then it is possible that the genotypes
will undergo gradual improvement until they are viable (highly 
fit). 
In our system, we thus ask for a steady state pattern ${\mathbf S}$
and subject the genotypes to a selection, for instance 
a selection proportional to
the fitness as given in Eq.~(\ref{eq:fitness}). 
If such selection is done on small populations,
the resulting dynamics is quite similar to what arises in the Metropolis
random walk algorithm, so we have implemented this \emph{in silico}
evolution process using the Monte Carlo algorithm described in the
previous section.

In effect, identifying each of the attempted changes in our algorithm
as a mutation, we examined the change in the fitness with
increasing number of mutations. 
\begin{figure}
\includegraphics[width=8cm]{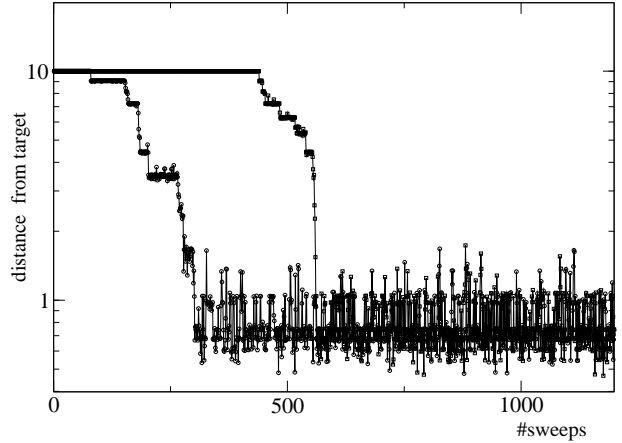}
\caption{\label{fig:History}  Building up
the stationary expression pattern ${\mathbf S}$ to ${\mathbf S}^{(target)}$,
starting with a random genotype. Shown are two typical trajectories as 
a function of the number of mutational sweeps. Here $N=20, L=12, 
\varepsilon=2.0$ and $h=0.01$.}
\end{figure}
The result of this simulation is shown in Fig.~\ref{fig:History}, where 
we plot the
fitness (actually the distance $D({\mathbf S},{\mathbf S}^{(target)})$) 
versus the number of mutational \emph{sweeps}
for two typical realizations. It is instructive to
follow such a trajectory in more detail. Let us rewrite $P_{ij}(t)$
(\emph{cf.} Eq.~(\ref{eq:dynamicsBis})) as
\begin{equation}
P_{ij}(t) = S_j(t)/(S_j(t) + x_{ij}) \; ,
\label{newP}
\end{equation}
\noindent
where $x_{ij}=h/W_{ij}$. In a random system, 
the $W_{ij}$'s are grouped around the average 
value 
\begin{equation}
\langle W_{ij}\rangle = (0.25+0.75 ~ e^{-\varepsilon})^L \ .
\label{averageW}
\end{equation}
When $L=12$ and $\varepsilon = 2$, this quantity is 
fairly small, about $3.6 \times 10^{-6}$. 
Hence, for $h > 10^{-4}$, all $x_{ij}$ have large values. Consequently, 
for the initial (random) genotypes, Eq.~(\ref{eq:dynamics}) yields 
$S_i \approx 0$ on the l.h.s., and this is of course
a steady state expression pattern. Since, 
by construction, ${\mathbf S}^{(target)}$ 
has $N/2$ elements equal to $1$, all other 
being $0$, the distance is 
then $D\approx N/2$. Such a situation may persist 
for a rather long time, 
until as the result of continuing mutations 
an unlikely event occurs: a diagonal 
element of the array $x_{ij}$, say $x_{kk}$, 
becomes small enough by chance. Then, one 
of the dynamical equations basically reduces to
\begin{equation}
S_k(t+1) \approx S_k(t)/(S_k(t) + x_{kk}) \; ,
\label{nontrivial}
\end{equation}
with a non-trivial stationary solution 
$S_k \approx 1 - x_{kk}$. 
For $x_{kk} =h \exp(\varepsilon d_{kk})$ to be $O(1)$ or smaller, 
$d_{kk}$ must be less 
than $d_h = [\ln{(1/h)}/\varepsilon]$, a
value we call ``critical'' hereafter. For 
the parameter range under consideration,
the generic result is that $S_k$ approaches values close to 1
and a relatively strong interaction appears. Such an interaction - which
will turn out to be ``essential'' as explained soon - is characterized by 
a subcritical mismatch. In this situation, the 
distance $D$ to the target phenotype drops by 
an amount of order $O(1)$. Notice, that a subsequent 
significant \emph{increase} of this distance is 
highly improbable because of the selection pressure: one unit of
increase is counter-selected by a factor $\exp(-\sqrt{n})$ from
what was explained in the previous section.

As the process continues with additional mutations, the increase
in fitness dramatically accelerates. This is because the
presence of every essential interaction has 
a double effect: it renders more probable the formation
of new essential interactions which no longer need 
to be located on the diagonal, and it tends
to lift up simultaneously several $S_i$'s. The distance to the target
phenotype then
rapidly drops step by step until the system reaches a regime
where $D$ fluctuates around a plateau value. For the choice
of parameter values used in Fig.~\ref{fig:History}, this leads to 
$S_i \approx 0.94$ for each $i$ satisfying
$S_i^{(target)} = 1$, while the remaining 
expression levels are negligibly small. From this ``evolution'' experiment,
we see that our model allows for evolvable genotypes, in the sense
that a new function can be acquired under realistic selection
pressures if given enough time.
\par
   It should be emphasized that a large auto-regulation $W_{kk}$
appears as the first noticeable evolutionary event (when a
diagonal mismatch becomes small enough by chance). While such
a direct auto-regulation is
necessary to initiate the turning on of genes
in our simulation of evolvability,
it is not crucial later. We have thus measured
the fraction of large auto-regulatory interactions
at long times (when the Monte Carlo provides
equilibrium samples of viable) to
see if such interactions remain favored. The
answer is no: the fraction
of auto-regulatory interactions is no larger than
expected by chance.
\par
We should also mention that once a viable genotype has been found,
in practice we find it to be viable also when using other
choices for ${\bf S}^{(initial)}$ other than ${\bf S}=0$. 
(Thus the convergence of Eqs.~(\ref{eq:dynamics}) and (\ref{eq:dynamicsBis})
to the fixed point is largely independent of the starting expression pattern.)
It seems that without introducing repressor interactions
(which is beyond the scope of this paper), it is not possible to 
condition viability on ${\bf S}^{(initial)}$ in our model.

\subsection{Sparse essential interactions}

Remaining in the framework of having the TF encodings fixed,
the Monte Carlo evolutionary dynamics 
just described will generate at large times
an equilibrium distribution of fit (viable) genotypes. Recall that, because of 
the choice of our
acceptance criterion, the algorithm will sample 
genotypes with a frequency proportional to their fitness. 
Hereafter we shall call this the space of viable genotypes.
Viable genotypes define in essence a null hypothesis since that
space determines the
distribution of any GRN property arising from the sole constraint
of function, \emph{i.e.}, the only constraint imposed on the
GRN is that they be fit. We will now see that the vast majority
of these GRN are in fact sparse, a feature that does not
at all arise in GRN models that ignore the microscopic
origin of the $W_{ij}$.

\begin{figure}
\includegraphics[width=8cm]{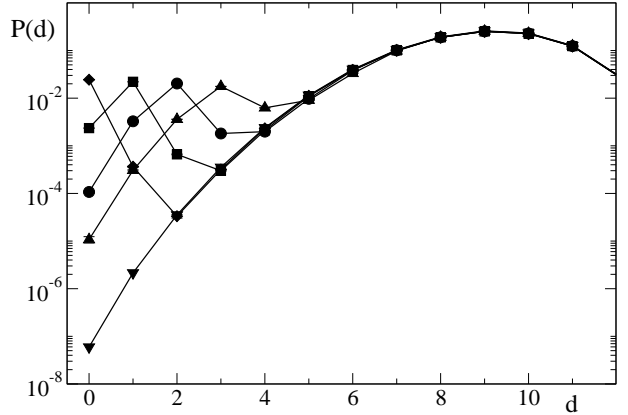}
\caption{\label{fig:mismatch} 
Distribution of the Hamming distance 
between a TF and the receiving DNA site 
for $N=20, L=12, \varepsilon=2.0$ and 
for various values of the threshold parameter:
(diamond) $h=0.1$, (square) $h=0.01$, 
(circle) $h=0.001$, (triangle up) $h=0.0001$, (triangle down) binomial. 
The lines are to guide the eye.}
\end{figure}

For our computational study, as suggested in Fig.~\ref{fig:History},
it is best to perform measurements every few hundred sweeps so that the
data collected are not overly correlated. We typically 
used $10000$ measurements separated by 100 sweeps. 
In Fig.~\ref{fig:mismatch} we show the distribution of the mismatch
$d$ when there are $N=20$ genes, for increasing values of
the ``threshold'' parameter $h$. At the low values of $h$,
$d$ has a binomial distribution with a peak
near $d=3L/4$ as expected. However at small $d$ one observes
significant deviations. In fact, as $h$ increases, 
the viability constraint becomes more marked and 
the distribution becomes bimodal: a second peak
appears at low mismatch values. Note that 
this peak shifts as $h$ increases, indicating
that there are nearly perfect matches that appear
in that regime. 

Two remarks are in order: for intermediate values of $h$ it 
may happen that the peak at small $d$ extends over two
neighbor values of $d$; the strong interactions then do
not necessarily have the same strength. (Note that because
the $d$ take on discrete values, so do the $W_{ij}$.)
Also, one could
argue that the fitness parameter $f$ should be scaled like
$\sqrt{h}$ when $h$ is changed. We have carried out
calculations with such a rescaling but this change
does not modify substantially the overall picture.

\begin{figure}
\includegraphics[width=8cm]{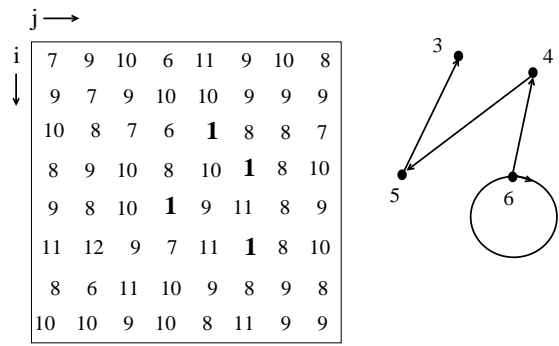}
\caption{\label{fig:example} 
Table of mismatches $d_{ij}$ between 
the $j$-th TF and its binding site in the regulatory region of 
gene $i$ for a toy GRN with $N=8$ (other parameters are as in 
Fig.~\ref{fig:History}). The mismatches corresponding to essential 
interactions are shown in boldface. In virtue of Eq.~(\ref{eq:Wij}),
the strength of the essential interactions is $\approx 0.135$
while the other interaction strengths are negligible 
(at most of order $10^{-6}$).
On the r.h.s. we have drawn the network, keeping essential interactions only.
The level of expression of the genes that are active in the target state is 
$\approx 0.926$ and that of the inactive ones ranges between $3\times 
10^{-6}$ and $6\times 10^{-4}$.}
\end{figure}

In Fig.~\ref{fig:example} we show an illustrative case of the mismatches
for a viable network with $N=8$, $\varepsilon = 2.0$ and $h=0.01$. 
The bimodal nature of the mismatch distribution is clearly apparent;
note in particular that the distribution at large $d$ is broad while
that at small $d$ is narrow. Recalling that
$W_{ij} = \exp(-{\varepsilon d_{ij}})$, the interaction matrix can be
considered to be sparse if we focus only on strong interactions.
However doing so requires defining rather arbitrarily a distinction
between strong and not strong $W_{ij}$. This difficulty is 
inherent to our framework where $W_{ij}$ is never 0, in contrast to
the situation in other gene network models where an interaction
is present or absent. 

\begin{figure}
\includegraphics[width=8cm, angle=0]{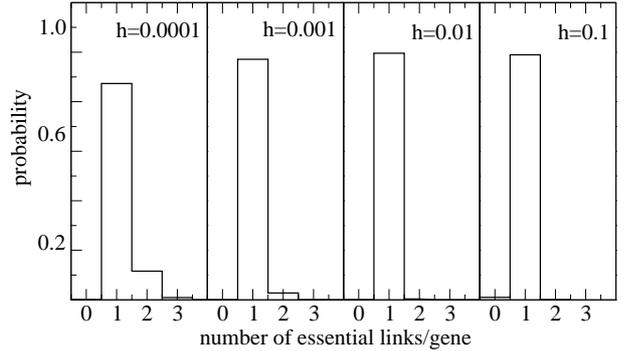}
\caption{\label{fig:essential}  Probability distribution 
of the number of essential interactions
per row of the matrix specifying a viable network for $N=20, L=12, 
\varepsilon=2.0$ and a range of values of $h$.}
\end{figure}

Since reality does not follow a ``present/absent''
dichotomy for interactions, it is desirable to find a way to 
define mathematically a notion of sparsity in our context without too 
much arbitrariness. For this, let us
consider not
the weights themselves but their functionality. For a given
genotype, call an interaction $W_{ij}$ ``essential'' if 
when setting it to zero viability is lost.
With this definition, one can ask whether viable genotypes
have sparse essential interactions, and in particular 
whether there are
few essential interactions per row of the matrix $\{W_{ij}\}$.
Notice that the number of active rows 
(\emph{i.e.}, having not too small expression
levels) must be equal to the number of genes that are ``on'' in the target
phenotype, $N/2$ in our case, otherwise the fitness is
far too low. We find that as soon as $h$ is not too small, 
there is almost always just one essential interaction per row as shown in 
Fig.~\ref{fig:essential} for $N=20$ and $L=12$. The same 
result holds for other relevant values of $N$, $n$ and $L$, suggesting that
within our model, the drive towards sparse interactions arises in 
regimes of biological relevance. 

We also considered a stronger measure
of essentiality: we asked that viability be lost when 
the interaction's mismatch is increased by one. Remarkably, the rule 
``one essential incoming interaction per gene'' generally held here too. Thus 
mutations in these interactions are typically deleterious, while 
mutations in the vast majority of the other interactions have
no consequence on the fitness. This shows that mutational
robustness is very heterogeneously distributed among the 
interactions in the network. The average robustness 
of fitness with respect to
binding site mutations, $R_{bs}$, is readily estimated: there are $N/2$ 
sensitive interactions out of $N^2$, hence one expects 
$R_{bs} \approx 1-1/2N$. And indeed we find $R_{bs} = 0.977(2)$ for $N=20$ 
(with some weak dependence on $h$ in the third decimal).

A consequence of the above is that 
in each active row of a viable network
the mismatch distribution is 
(semi-quantitatively) well represented 
by the ansatz ($H(x)=1$ for $x>0$ and $0$ otherwise): 
\begin{equation}
\tilde{p}(d) \approx
\frac{N-1}{N} p(d) + \frac{1}{N} \frac{p(d) H(d_h - d)}{\sum_{m<d_h}
  p(m)} 
\label{eq:marginal}
\end{equation}
with a very simple interpretation:  
the shape of the probability distribution of any mismatch 
is essentially that without the viability constraint, but with
an additional peak at small values of the mismatch. 
There is thus one ``leading'' mismatch taking 
care of most of the viability constraint, while 
the other mismatches behave approximately 
as if they were unconstrained.
\par
It is interesting to see how the 
essential interactions are distributed 
among the columns of $\{W_{ij}\}$. First 
of all, they may appear in $N/2$ columns only
(in our example). This is an immediate 
consequence of the target phenotype
being a fixed point of the dynamical process 
Eq.~(\ref{eq:dynamics}). A simulation yields
the result that, as one might expect, 
the $N/2$ essential interactions are distributed
completely randomly among the $N/2$ active 
columns. Since the number of essential
interactions in a column is a measure of the 
effective out-degree of the corresponding node of
the GRN (in the theory of weighted networks 
one would rather refer to the {\em strength} 
of the network node), one readily convinces 
oneself that the out-degree distribution
is binomial with probability parameter equal to $2/N$.
More generally, let $K$ be the number of active
genes in ${\bf S}^{(target)}$. Viable GRN will typically
have $K$ essential reactions, all connecting these
active genes. Furthermore, these essential interactions 
define a subnetwork where the in-degree is almost always
1 while the out-degree has a binomial distribution that is
well approximated by a Poisson law of mean 1.

A long succession of binding site mutations 
can move an essential interaction from
one column to another. A few hundred sweeps 
are sufficient for that. Measuring the 
moments of the out-degree distribution 
one can calculate the autocorrelation time 
of this process. It turns out to be 
roughly of the order of $10^3$ sweeps for $h=0.01$
and about $10^{2}$ sweeps for $h=0.001$. 
This implies that very different
genotypes can arise under mutation-selection 
balance if given enough time;
our GRN model thus allows for gradual change of genotypes while
maintaining phenotypes, a characteristic of evolvability.

\section{TF mutations and a population dynamics approximation}

Up to now we assumed that the strings 
of characters associated with TFs 
were fixed. Of course, the genes coding 
for TFs can mutate. These mutations
have very specific consequences for the 
GRN: when a TF is modified in our
model, a whole column
of the matrix $\{W_{ij}\}$ is affected
at once. There will thus be 
a selective pressure on networks depending
on how their essential interactions are
spread across different columns, and this
will change the out-degree distribution.



Qualitatively, one expects broad distributions to be favored; indeed, consider
mutating a TF coding gene. If the 
TF is associated with at least one essential interaction, then such a
mutation is typically highly deleterious. Oppositely, if the TF has no 
essential interactions, the mutations will be 
innocuous. Hence, in the presence of TF mutations, there is a strong
correlation between the robustness $R_{tf}$ to mutations in the 
different TFs and column ``occupancy'' by essential 
interactions (\emph{i.e.}, the essential out-degree, to 
be denoted $k_{out}$). Let $K$
be the total number of essential interactions 
(typically equal to $N/2$ in the results presented here), 
and let $N_{occupied}$ ($N_{empty}$) be
the number of occupied (empty) columns. Of 
course $N_{occupied} + N_{empty} = N$.
Since mutating essential interactions is highly deleterious,
the average robustness with respect to mutating a TF coding gene is
accurately given by
$R_{tf} = N_{empty}/N$ and it is easy to convince oneself that 
$\langle k_{out}\rangle = K/N_{occupied}$ 
where the average $\langle \cdot \rangle$
is to be taken only over occupied collumns. 
After elementary algebra one gets
\begin{equation}
R_{tf} = 1 - \frac{K}{N\langle k_{out} \rangle} \; .
\label{TFrobGeneral}
\end{equation}
In the examples discussed in this paper, $K=N/2$, and therefore
\begin{equation}
R_{tf} = 1 - \frac{1}{2 \langle k_{out} \rangle} \; ,
\label{TFrobustness}
\end{equation}
a result fully confirmed by an explicit 
numerical simulation. Thus, GRN with smaller
$k_{out}$ are disfavoured as soon as we allow TFs to mutate.

To reveal the effects of selection, we must 
work with a population of GRN under 
mutation-selection balance and see the networks with broad
out-degree distributions out-compete those with more 
narrow distributions.
A large scale simulation of such a
population would provide the distribution of
$k_{out}$. Here we choose a simplified approach based
on \emph{effective} evolutionary dynamics as 
follows. First, one can safely ignore
all the inactive genes in ${\mathbf S}^{(target)}$, and we are left with
$K$ different TFs. Second, for each of these $K$ genes, there will almost always 
be exactly one essential (incoming) interaction. We thus map each
microscopic genotype to an effective genotype, specifying only the
essential interactions, as can be justified from the decoupling
of Eq.~(\ref{eq:marginal}). Third, we consider a population of such
effective genotypes undergoing mutation and selection. At the microscopic
level, we showed in the previous section that mutations 
of the binding sites will lead to 
slow changes in the positions of the essential interactions.
These changes arise at the time scale $\tau$ which empirically we found to be
several hundred sweeps, each sweep corresponding to $L N^2$ mutations.
That time scale $\tau$ is very large for the population
dynamics under consideration so for simplicity think of 
$1/\tau$ as being infinitesimal. 
Forth, at each generation we evolve the population 
of (effective) genotypes:
the individuals are in competition
because mutations in the TF coding genes affect them differently.
The relevant characteristic of each genotype is its 
mutational robustness $R_{tf}$ to mutations in these genes.
Thus $R_{tf}$ acts as a fitness which is under selection,
with large $R_{tf}$ being favored; in view of 
Eq.~(\ref{TFrobustness}), this should tend
to enhance large out-degrees.

To see how this transpires within a population genetics framework,
we consider an effective model where an
essential interaction is represented by a ``ball'',
a column of $\{W_{ij}\}$ by a ``box'' and an individual
(e.g. a cell) in the population by $K$ boxes. The population of such
``cells'' is studied numerically simulating the
well-known Moran process \cite{MoranBook1962} as follows.
At each step, we take a random individual of the population; let
$N_{occupied}$ be its number of genes with out-degree greater or
equal to 1. A selection process is applied to this individual.
With probability $R_{tf} = 1-N_{occupied}/K$ it is duplicated
and another randomly chosen individual is removed from the population.
Otherwise (and thus with probability $N_{occupied}/K$),
one tries another random individual,
and so forth. With this Moran process, the population size stays constant
as the fitness rises and reaches a limit.
Higher fitness (robustness $R_{tf}$) corresponds via
Eq.~(\ref{TFrobustness}) to
a broader out-degree distribution, large degrees being favored.

\begin{figure}
\includegraphics[width=8cm]{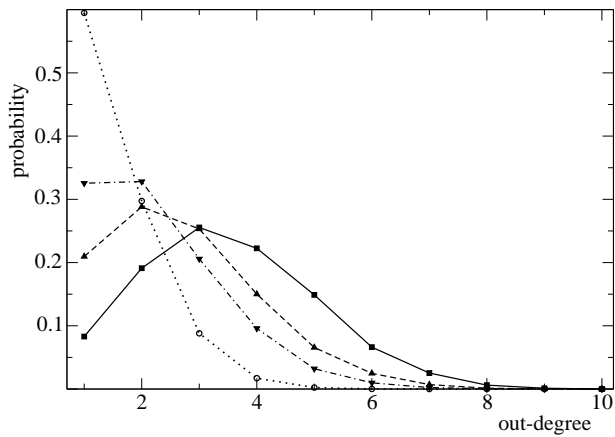}
\caption{\label{fig:outDegree} Out-degree distribution within the simplified
genotypes (see text) when using the Moran process. The population consists of
$100000$ genotypes and $K=10$: squares,
$K=20$: triangles up, $K=40$: triangles down (circles refer to the binomial 
distribution of mean 1, which for these values of 
$K$ is almost $K$-independent).}
\end{figure}

The simulation of this simplified population model
is much faster than if we were to use the microscopic genotypes.
It is also far more intuitive: each effective genotype can be
thought of as a way to put $K$ balls into $K$ boxes.
Each box is a TF and the balls in that box give the
number of essential interactions associated with that TF.
In Fig.~\ref{fig:outDegree} we display 
the resulting out-degree distribution for $K = 10, 20$ and $40$.
The initial condition on the simulation is that the 
balls are distributed at random among the $K$ boxes.
At this point a few comments are in order:

- For the choice $1/\tau = 0$, the result 
of the Moran process may depend on the initial 
condition. Since, in our case,
the distribution of essential interactions 
among columns of $\{W_{ij}\}$ generated by binding
site and TF mutations is different, one may 
worry that the results shown in Fig.~\ref{fig:outDegree}
change if one mixes individuals with binomial 
and flat ball distributions.  We have found that mixing 
them in equal proportion leads to an out-degree 
distribution that has almost the same shape as
in Fig.~\ref{fig:outDegree}, except that 
it is shifted to the left by about 0.5, a minor modification.

 - In reality, $\tau$ is not infinite, so at each generation, 
one should allow, at the rate
$1/\tau$, balls to go from one box to another. 
We checked explicitly that hops 
of balls done with frequency 1/10000, still much larger than is realistic,
do not change the result for the out-degree distribution in a perceptible way. 

\section{Discussion and conclusion}

We considered a quite simple model of a gene regulatory network (GRN)
in which function is identified with reaching 
given target gene expression levels.
The key feature which sets our framework apart is the introduction
of molecular genotypes which thereby specify
interaction weights in the GRN. Within this setting,
we have shown how a ``viable''(\emph{i.e.}, having the desired function) GRN
is progressively constructed under selection pressure, when one forces the
phenotype to increase its ``fitness'' by approaching a previously defined
optimal phenotype. We investigated the properties of networks in this model 
under the constraint that they be viable. We find that for 
a certain range of values of the model's parameters, 
the viability constraint leads to \emph{sparse} GRN; 
we have quantified this through the 
sparsity of ``essential'' interactions. 
Interestingly, the effects of the viability 
constraint condense onto just a few of 
the interactions, the others 
being non-functional. As a result, nearly all mutations of the binding sites 
have no effect on the viability and so such sites have a very high mutational 
robustness. However, for those few sites which 
bear the essential interactions,
the majority of mutations are deleterious 
so their mutational robustness is low. 
Thus in our GRN, the mutational robustness 
is extremely heterogeneous from site 
to site. In addition, any ``redundant'' 
interaction is expected to become lost 
under evolutionary dynamics since mutations 
will remove it and condense the 
burden of viability onto a smaller number of interactions. We have also
studied the consequences of TF mutations for the fate of a population.
In contrast to the behavior of the in-degree distribution
which is ``as narrow as possible'', we find that the 
distribution of out-degrees develops a fat tail.

Although our modeling involves certain 
idealizations, its main characteristics 
are fairly realistic; in particular we 
have insisted on including interactions 
through the biophysical mechanism of molecular 
recognition and affinity. It is 
therefore quite striking that a reasonable 
GRN topology comes out very naturally 
in this framework. It should be clear that this 
success is a result of the combined 
effect of several causes: the viability 
constraint, the low probability of a small 
mismatch between TF and the binding site on the DNA, 
the size $L$ of this segment, the 
not-too-small spacing (in units of $k_B T$) 
between the energy levels that determine 
the strength of TF-DNA interactions, and finally the value 
of the threshold parameter $h$ itself which enters the dynamics 
of the gene expression levels (\emph{cf.} Eq.~(\ref{eq:dynamics})).

\begin{figure}
\includegraphics[width=8cm]{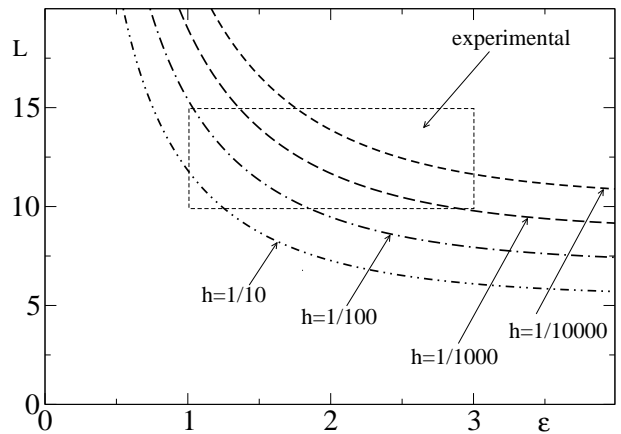}
\caption{\label{fig:phaseDiagram}  The curves show the lower 
limit of the region where 
$h = 10 N [0.25+0.75 exp(-\varepsilon)]^L$ when $N=20$.  }
\end{figure}

This overall picture corresponds to having all 
factors $P_{ij}$ in Eq.~(\ref{eq:dynamics}) be rather small except
for the one which is associated with the essential incoming interaction.
Such a scenario arises when the threshold $h$ 
is significantly larger than the sum of 
random $W_{ij}$ entering these factors (behaving, 
in practice, as if the viability constraint 
were absent). For a four letter code, assuming, 
as we do, that the binding energy is additive 
in mismatches and that every mismatch costs the same, one gets the condition
\begin{equation}
h \gg N (0.25+0.75 e^{-\varepsilon})^L
\label{limits}
\end{equation}
Note that within a two letter code, the condition forces 
one to larger values of $L$ (close to 20) and thus 
beyond what is realistic biologically.
Since the model parameters correspond to 
measurable quanties, it is appropriate to compare to 
biological values. According to Eq.~(\ref{eq:probOccupy})
the probability that a TF occupies a DNA site
is controlled by $h=1/n$, where $n$ is the number of these 
molecules. We considered the range 
$1/10000 < h < 1/10$. Interpreting $\gg$ as
``larger by one order of magnitude'', \emph{i.e.}, by
a factor 10, one gets an allowed region in parameter
space as illustrated in Fig.~\ref{fig:phaseDiagram}.

This figure can be used to obtain the predicted domain of relevance:
it is above the corresponding curves (taken at $N=20$ and illustrative values
of $h$). We see that $\varepsilon$ and $L$ should not be
too small. Moreover, it is gratifying that the experimental 
range of these parameters (indicated by a rectangle) is near 
the border and, most of it, within this region. The model 
would remain meaningful if $L$ and $\varepsilon$ were even 
larger. However, in the analogue of Fig.~\ref{fig:mismatch}, 
the point at $d=0$ would dominate strongly over the few neighboring
$d$ points, and the GRN would be robust but not as
evolvable~\cite{BergWillmann2004,MustonenKinney2008}.
It is worth emphasizing that as the number $N$ of genes grows, 
it is necessary to increase slowly either $L$, $\varepsilon$ or $h$.
$\varepsilon$ is constrained by biophysical processes and thus not
evolvable, and $L$ seems the best candidate for the system
to adapt to increasing $N$~\cite{SenguptaDjordjevic2002}. Note 
nevertheless that the effects of growing $N$ are mild and that 
in practice regulation is modular, so effectively biological GRN
have only modest values of $N$.

\subsection*{Acknowledgments}
We thank V. Hakim, the late P. Slonimski and A. Wagner for helpful comments.
This work was supported
by the EEC's FP6 Marie Curie RTN under
contract MRTN-CT-2004-005616 (ENRAGE: European
Network on Random Geometry), 
by the EEC's IST project GENNETEC - 034952
and by the Polish Ministry of Science
Grant No.~N~N202~229137 (2009-2012). 
Project operated within the Foundation for Polish Science
International Ph.D. Projects Programme co-financed by the European Regional
Development Fund covering, under the agreement no. MPD/2009/6, the
Jagiellonian University International Ph.D. Studies in Physics of Complex
Systems.
The LPT and LPTMS are Unit\'e de Recherche de
l'Universit\'e Paris-Sud associ\'ees au CNRS.


\bibliographystyle{apsrev}
\bibliography{grn}


\end{document}